\documentclass[twocolumn]{article}

\usepackage{graphicx}

\newcommand{\be}{\begin{equation}}
\newcommand{\ee}{\end{equation}}
\newcommand{\kk}[2]{\frac{#1}{#2}}
\newcommand{\bm}[1]{\mbox{\boldmath$ #1 $}}
\newcommand{\pab}[2]{\frac{\p #1}{\p #2}}
\newcommand{\pcd}[2]{\frac{\p^2 #1}{\p #2^2}}

\def\p{\partial}
\def\s{\,\,\,\,}
\def\0{^{(0)}}
\def\1{^{(1)}}
\def\n{\eta}
\def\={\approx}

\def\z{\zeta}
\def\v{\lambda}
\def\ra{\rightarrow}

\title{\Large \bf Nonlinear Viscoelastic Compaction in Sedimentary Basins }
\author{Xin-She Yang \\
  {\small Department of Fuel and Energy and Applied Mathematics,
  University of Leeds, LEEDS LS2 9JT, UK }}
  
\date{}
\begin{document}
\maketitle

\begin{abstract}
In the mathematical modelling of sediment compaction and porous media flow,
the rheological behaviour of sediments is typically modelled in terms of a
nonlinear relationship between effective pressure $p_e$ and porosity $\phi$,
that is $p_e=p_e(\phi)$.  The compaction law is essentially a
poroelastic one. However,  viscous compaction
due to pressure solution becomes important at larger depths
and causes this relationship to become more akin to a
viscous rheology. A generalised viscoelastic compaction model of Maxwell
type is formulated, and different styles of nonlinear behaviour
are asymptotically analysed and compared in this paper.  \\
\end{abstract}

\noindent {\sf Citation detail}: X. S. Yang, Nonlinear viscoelastic compaction in 
sedimentary basins, {\it Nonlinear Processes in Geophysics}, {\bf 7}, 1-7 (2000).

\section{Introduction}

Drilling mud  is used in well-bores  drilled for oil exploration
to maintain the integrity and safety of the hole.
The mud density must be sufficient to prevent collapse of the hole, but not
so high that hydrofracturing of the surrounding rock occurs. Both these
effects depend on the pore fluid pressure in the rock, and drilling
problems occur in regions where abnormal pore pressure or {\em
overpressuring} occurs, such as in the
sedimentary basins of the North Sea,  where pore pressure
increases downward faster than hydrostatic pressure.
Such overpressuring can affect oil-drilling rates
substantially and even cause serious blowouts during drilling.
Therefore, an
industrially important objective is to predict overpressuring
before drilling and to identify its precursors during drilling.
Another related objective is to predict reservoir quality and
hydrocarbon migration. An essential step in the achievement of
such objectives
is the scientific understanding of their mechanisms and the
evolutionary history of post-depositional sediments such as  shales.

Shales and other fine-grained compressible rocks are considered to be
the source rocks for much petroleum found in sandstones and carbonates.
At deposition, sediments such as shales and sands
typically have  porosities of about $0.5 \sim 0.75$  (Lerche, 1990).
When sediments are drilled at a depth of say 5000 m, porosities are
typically $0.05 \sim 0.2$. Thus an enormous
amount of water has escaped from the sediments during their deposition
and later evolution. Because of the fluid escape, the grain-to-grain
contact pressure must increase to support the overlying sediment
weight. Dynamical fluid escape depends lithologically on the permeability
behavior of the evolving sediments. As fluid escape proceeds, the porosity
decreases, so the permeability becomes smaller, leading to an ever-increasing
delay  in extracting the residual fluids. The addition of overburden
sediments is compensated by an increase of excess pressure
in the retained fluids. Thus overpressure develops from such a
{\em non-equilibrium compaction} environment (Audet and Fowler, 1992;
Fowler and Yang, 1998).
A rapidly accumulating basin is unable to expel pore fluids sufficiently
rapidly due to the weight of overburden rock. The development of
overpressuring retards compaction, resulting in a higher porosity, a higher
permeability and a higher thermal conductivity than are normal for a
given depth, which in turn changes the structural and stratigraphic shaping of
sedimentary units and provides a potential for hydrocarbon migration.
Therefore, the
determination of the mechanism of the dynamical evolution of escape
of fluids
and the timing of oil and gas migration out of such fine-grained rocks
is a major problem. The fundamental understanding  of mechanical and
physico-chemical properties of  these rocks in the earth's crust
has important applications in petrology, sedimentology, soil mechanics,
oil and gas engineering and other geophysical research areas.

Compaction is the process of volume reduction via
pore-water expulsion within sediments due to the increasing weight of
overburden load. The requirement of its occurrence is not only the
application of an overburden load but also the expulsion of pore water.
The extent of compaction is strongly influenced by the burial history and
the lithology of sediments. The freshly deposited loosely packed
sediments tend to evolve as an open system, toward a closely packed
grain framework during the initial stages of burial compaction and this is
accomplished by the processes of grain slippage, rotation, bending and
brittle fracturing. Such reorientation processes are collectively
referred to as {\em mechanical compaction},
which generally takes place in  the first 1 - 2 km of burial.
After this initial porosity loss,
further porosity reduction is accomplished by the process of
{\em chemical compaction} such as pressure solution  (Fowler and Yang,
1999).

Despite the importance of compaction and diagenesis for geological
problems, the literature of quantitative modelling is not a huge one,
though the processes have received much attention in the literature,
and the mechanism leading to pressure solution is still poorly understood.
The effect of
gravitational compaction was reviewed by Hedberg (1936) who suggested that
an interdisciplinary study involving soil mechanics, geochemistry,
geophysics and geology is needed for a full understanding of the
gravitational compaction process.  More comprehensive and
recent reviews on the subject of compaction of argillaceous
sediments were made by Rieke and Chilingarian (1974) and
Fowler and Yang (1998).

Compaction is a density-driven flow in a porous medium,
a fascinating multidisciplinary  topic that has attracted  attention
from scientists with different expertise for a long  time.
Holzbecher (1998) provides an up-to-date comprehensive review
of the previous work and state-of-art  numerical methods and
softwares for modeling density-driven flow and transport in porous media
where the constant porosity is used. Here, we will mainly model
how porosity changes with time and depth, rather than using
a constant density; thus an appropriate compaction relation is
vitally important.

Nonlinear compaction models have been formulated in two ways.
The early and most models used elastic or
poroelastic rheology, and the compaction relation is of Athy's type $p_e=p_e
(\phi)$    (Gibson, England \& Hussey, 1967; Smith, 1971;
Sharp, 1976; Wangen, 1992; Audet and Fowler, 1992; Fowler and Yang, 1998).
More recent models use a viscous rheology with
a compaction relation $p_e=-\xi \nabla . {\bf u}^s$ (Angevine and Turcotte,
1983; Birchwood and Turcotte, 1994;  Fowler and Yang, 1999).
The poroelastic models are valid for mechanical compaction while
the viscous models mainly describe the chemical compaction such as
pressure solutions.
More recently, efforts have been made to give a  more realistic
visco-elastic model (Revil, 1999; Fowler and Yang, 1999).
Fowler and Yang (1999) use a viscous law $p_e=-\xi \nabla . {\bf u}^s$
to model compaction due to pressure solution, while Revil (1999) uses
a poro-visco-plastic model, with a relationship between  porosity
strain and effective stress, to study pressure
solution mechanism and its applications. However, there is no viscoelastic
model which has been formulated to analyse the compaction problem on a basin
scale, and most of the conventional studies are mainly numerical simulations.
Obviously more work has yet to be done. This paper aims at providing a
unified approach to the compaction relation
by using a visco-poroelastic relation of Maxwell type. The nonlinear partial
differential equations are then analysed by using asymptotic methods and the
analytical solutions are compared with numerical simulations.

\section{Mathematical Model}

For the convenience of investigating the effect of sediment compaction,
 we will assume a single species only.   The sediments \, act as \, a
compressible porous matrix, \, so that mass conservation of \, pore fluid
together  with Darcy's law leads to the equations:

\noindent {Conservation of mass,}
\begin{equation}
\frac{\partial}{\partial t}(1-\phi)+{\bf \nabla} \cdot
[(1-\phi) {\bf u}^{s}]=0,  \label{VC:MASS}
\end{equation}
\begin{equation}
\frac{\partial \phi}{\partial t}+{\bf \nabla} \cdot (\phi {\bf
u}^{l})=0,
\end{equation}
{Darcy's law,}
\begin{equation}
\phi ({\bf u}^{l}-{\bf u}^{s})=-\frac{k}{\mu}(\nabla p^{l}+\rho_{l} g {\bf j}),
\end{equation}
 {Force balance,}
\begin{equation}
\nabla \cdot \mbox{\boldmath $\sigma^{e}$ }-\nabla [{p}^{l}]-\rho
g {\bf j}={\bf 0},
\end{equation}
where  ${\bf u}^{l} $ and ${\bf u}^{s}$ are the
velocities of the fluid and solid matrix, $k$ and $\mu$ are the matrix
permeability and the liquid
viscosity,   $\rho_{l} $ and $\rho_{s}$ are the
densities of the fluid and solid matrix,   $\mbox{\boldmath $\sigma_{e}$ }$ is
the effective stress, $p_{e}$ is the effective pressure,
${\bf j}$ is the unit vector pointing vertically upwards,
$p^{l}$ is the pore pressure, and $g$ is the gravitational acceleration.
In addition, a rheological  compaction law is needed to complete this
model.

\subsection{Poroelasticity and Viscous Compaction}

The compaction law is a relationship between effective pressure $p_e$ and
strain rate $\dot e$ or porosity $\phi$. The common approach in soil
mechanics and sediment compaction is to model this generally nonlinear
behaviour as  poroelastic, that is to say, a relationship of Athy's law
of the form
$p_e = p_e (\phi)$, which is derived by fitting real data for sediments.
Athy's poroelasticity law is also a simplified form of Critical State Theory
(Schofield and Wroth, 1986).
A common relation describing  poroelasticity is
\begin{equation}
\frac{D p_e}{D t}=-K \nabla \cdot {\bf u}^s,   \,\,\,\,\, \frac{D}{D t}=
\frac{\partial}{\partial t} + {\bf u}^s \cdot \nabla,
\end{equation}
and equation (\ref{VC:MASS}) can be rewritten as
\begin{equation}
\frac{1}{1-\phi} \frac{D (1-\phi)}{D t} = - \nabla \cdot {\bf u}^s,
\label{us-phi}
\end{equation}
combining with the previous equation, we have
\begin{equation}
p_e = p_e (\phi),
\end{equation}
which is Athy's law for poroelasticity. A typical form of
this constitutive relation (Smith, 1971; Audet and Fowler, 1992;
Fowler and Yang, 1998) is
\be
p_e= \ln (\phi_0/\phi)-(\phi_0-\phi). \label{pe-phi}
\ee
However, this poroelastic compaction law is  only valid for
sediment compaction
in the upper and shallow region, where compaction  occurs due to purely
mechanical movements such as  grain sliding and packing rearrangement. In the
deeper region, mechanical compaction is gradually replaced by
chemical compaction due to stress-enhanced flow along grain boundaries
from the grain contact areas to the free pore, where the pressure
is essentially the
pore pressure. A typical process of such chemical compaction in sediment
is pressure solution, whose rheological behavior is usually viscous, so that
it is sometimes called viscous pressure solution or viscous creep.

The mathematical formulation of  compaction laws for pressure solution
is to derive the creep rate in terms of concentrations, grain size and
geometry (usually spherical or cylindrical packings), effective
stress, grain boundary
diffusion. This allows us to include the detailed
reaction-transport process in a  relation between
strain rate and effective stress, although further simplifications
are usually assumed such as steady-state dissolution and
{\em local} reprecipitation along the grain boundary. Rutter's {\em creep law}
(1976) is widely used
\begin{equation}
\dot e=\frac{A_{k} c_{0} \, w D_{gb} }{\rho_{s} \bar d^{3}}
\sigma, \label{CREEP-1}
\end{equation}
where $\sigma$ is the effective normal stress across the grain contacts,
$A_{k}$ is a constant, $c_{0}$ is the equilibrium concentration
(of quartz) in pore fluid,  $\rho, \, \bar d$ are the density
and (averaged) grain diameter (of quartz). $D_{gb}$ is the
diffusivity of the solute in water along grain boundaries with a
thickness $w$.
Note that $p_{e}=-\sigma$ and $\dot e_{kk}={\bf \nabla} \cdot {\bf u^{s}}$.
With this, (\ref{CREEP-1}) becomes the following  compaction law
\begin{equation}
p_{e}=-\xi \nabla . {\bf u}^{s}.   \label{VC:CREEPNEW}
\end{equation}
This was first used by Birchwood and  Turcotte (1994) to study pressure
solution in sedimentary basins by presenting a unified
approach to geopressuring, low  permeability zone formation and
secondary porosity generation.

\subsection{1-D Viscoelastic Compaction}

Following the discussions of elastic compaction (Fowler and Yang
1998) and viscous compaction (Fowler and Yang, 1999), we can
generalise the above relations to a viscoelastic compaction law of
Maxwell type
\be
{\bm\nabla}.{\bf
u}^s=-\frac{1}{K}\frac{D p_e}{Dt}-\frac{1}{\xi}p_e.
\ee
Equivalently, we would anticipate that a viscoelastic rheology
holds for the medium,
involving material derivatives of tensors, and some care is needed to ensure
that the resulting stress-strain relation be  invariant
under the coordinate transformation.
This is not always guaranteed due to the complexity of the rheological
relations (Bird, Armstrong \& Hassager 1977). Fortunately, for
one-dimensional flow, which is always {\em irrotational},
the equation is invariant and all the different equations in corotational
and codeformational frames degenerate into the same form (Yang, 1997).
In the one-dimensional case we discuss below, we can take this for
granted. The 1-D model equations become
\begin{equation}
\pab{ (1-\phi)}{t} +\pab{}{z}[ (1-\phi) u^{s} ]=0,
\label{VC:MASS}
\end{equation}
\begin{equation}
\pab{ \phi}{t}+ \pab{( \phi  u^{l})}{z}=0,
\end{equation}
\begin{equation}
\phi (u^{l}-u^{s})
=\frac{k(\phi)}{\mu}[-G \pab{p_e}{z}-(\rho_s-\rho_l)(1-\phi) g ],
\end{equation}
\be
\pab{u^s}{z}=-\frac{1}{K}\frac{D p_e}{Dt}-\frac{1}{\xi}p_e,
\s \frac{D}{Dt}=\pab{}{t}+u^s \pab{}{z},
\ee
where $G=1+4 \eta/3 \xi$ is a constant describing the properties
of the sediments, and
$\eta$ is the viscosity of the medium.
By assuming that the densities
$\rho_s$ and $\rho_l$ are constants, we can see that only the density
difference $\rho_s-\rho_l$ is important to the flow evolution. Thus,
the compactional flow is essentially density-driven flow in a porous
medium (Holzbecher, 1998).

\section{Non-dimensionalization}

Let the length-scale $d$ is defined by
\begin{equation}
d=\{\kk{\xi \dot m_s G}{(\rho_s-\rho_l) g} \}^{1/2},
\ee
so that the dimensionless pressure is
$p=G p_{e}/(\rho_{s}-\rho_{l}) g d=O(1)$. Meanwhile,
we  scale $z$ by $d$, $u^{s}$ by
${\dot m}_{s}$, $t$ by $ d/{\dot m}_{s}$, permeability $k$ by
$k_{0}$. By writing $k(\phi)=k_{0} k^*$, $z=d z^*$, ..., and dropping
the asterisks, we have
\begin{equation}
-\pab{\phi}{t} +\pab{}{z}[(1-\phi) u^{s}]=0,
\label{VC:MASS-1}
\end{equation}
\begin{equation}
\pab{\phi}{t}+ \pab{(\phi u^{l})}{z}=0, \label{PHI:U}
\end{equation}
\begin{equation}
\phi (u^{l}-u^{s})=\v k(\phi) [-\pab{p}{z}-(1-\phi) ],
\end{equation}
\be
\pab{u^s}{z}=-\kk{\phi}{(1-\phi)^2}\kk{ D p}{Dt}-p,
\s \kk{D}{D t}=\pab{}{t} + u^s \pab{}{z},
\ee
where
\begin{equation}
\lambda=\frac{k_{0} (\rho_{s}-\rho_{l}) g} {\mu {\dot m}_{s}}.
\ee
In the above derivation, we have used the requirement that the
poroelastic case (\ref{pe-phi}) result in the limit as the
viscous rheology vanishes.

Adding (\ref{VC:MASS-1}) and (\ref{PHI:U}) and integrating, we have
\be
u^s=-\phi (u^l-u^s),
\ee
where $u=\phi (u^l-u^s)$ is the Darcy flow velocity. By using
the equation (\ref{us-phi}), we have
\begin{equation}
\pab{\phi}{t} =\pab{}{z}[(1-\phi) u^s],
\label{equ-1}
\end{equation}
\begin{equation}
u^s=\v (\kk{\phi}{\phi_0})^m [-\pab{p}{z}-(1-\phi) ].
\label{equ-2}
\end{equation}
\be
\kk{1}{(1-\phi)} \kk{D \phi}{D t}
=-\kk{\phi}{(1-\phi)^2}\kk{ D p}{Dt}- p.
\label{equ-3}
\ee
The constitutive relation for permeability $k(\phi)$ is nonlinear, and
of the form,
\be
k(\phi)=(\kk{\phi}{\phi_0})^m, \s m=8.
\ee
The boundary conditions at $z=0$ are
\be
\pab{p}{z}-(1-\phi)=0 \s ({\rm or \,\, equivalently,} \s u^s=0),
\label{bbb-1}
\ee
\be
\phi=\phi_{0}, \,\,\,p=0,
\ee
\be
\dot h=\dot m(t) + \v (\kk{\phi}{\phi_0})^m [\pab{p}{z}-(1-\phi)]
\s {\rm at} \s z=h(t),
\label{bbb-2}
\end{equation}
which is a moving boundary problem.

We estimate these parameters  by using values taken
from observations. From the typical values
of $\rho_{l} \sim 10^{3} \, {\rm kg\, m}^{-3}$,
$\rho_{s} \sim 2.5 \times 10^{3} \, {\rm kg\, m}^{-3},\,$
\,\,\,$ k_{0} \sim 10^{-15} -\!\!- 10^{-20}\, {\rm m}^{2}$,
$\mu \sim 10^{-3}\, {\rm N\,s\,
m}^{2}$, $\xi  \sim 1 \times 10^{21}$ N s $ {\rm m}^{-2}, $
$\dot m_{s}  \sim 300\, {\rm m\,\, Ma}^{-1}$
$=1 \times 10^{-11}\,
{\rm m\,\, s}^{-1},\, g \sim 10 {\rm m \,s}^{-2}, \, G \sim 1, \,
d \sim 1000 {\rm m}$ we get
$\v \= 0.01 \sim 1000$. The only
parameter $\v$, which governs the
evolution of the fluid flow and porosity in sedimentary basins,
is the ratio between the permeability and the sedimentation rate.

\section{Asymptotic Analysis}

Since the nondimensional parameter $\v \=0.01 \sim 1000$ that
controls  the  compaction process varies greatly,
we can expect that the two
limits $\v \ll 1$ and $\v \gg 1$ will have very different
features of porosity and flow evolution.  In fact,
$\lambda=1$ defines a transition between slow
compaction ($\lambda << 1$) and fast compaction ($\lambda >> 1$).
We follow the asymptotic analysis of Fowler and Yang (1998) and
Fowler and Yang (1999) to obtain some analytical asymptotic solutions.

\subsection{Slow Compaction ($\v \ll 1$)}

If $\v \ll 1$, $z \sim 1$, $t \sim 1$, $p \sim 1$, then
$u^s \ll 1$ and $\pab{\phi}{t} \= 0$, and it follows that
$\phi \= \phi_0$ and $D\phi/Dt \=\pab{\phi}{t}$. Thus
\be
\pab{\phi}{t} \= -\v (1-\phi_0) \pcd{p}{z},
\ee
\be
u^s \=\v [-\pab{p}{z}-(1-\phi_0)],
\ee
\be
\kk{1}{(1-\phi_0)} \pab{\phi}{t} \=-\kk{\phi_0}{(1-\phi_0)^2} \pab{p}{t}
- p,
\ee
which can be rewritten approximately as
\be
\pab{p}{t}=\v' \pcd{p}{z}+\kk{(1-\phi_0)^2}{\phi_0} p, \s \v'=\kk{(1-\phi_0)^2 \v}{\phi_0},
\ee
with the boundary conditions
\be
\pab{p}{z} \= 1-\phi_0, \s {\rm on} \s z=0,
\ee
\be
p \ra 0, \s z \ra \infty,
\ee
This is in fact equivalent to the case
of conduction in a semi-infinite space with a constant flux at $z=0$.
The solution in this case can be approximately expressed as
\begin{equation}
p \=(1-\phi_{0}) \sqrt{4 \lambda' t}\,\,
{\rm ierfc}(\z)+\sqrt{\v' \phi_0} \exp[-\kk{(1-\phi_0)z}{\sqrt{\v' \phi_0}}],
\ee
where
\be
\z=\kk{z}{\sqrt{4 \lambda' t}},
\end{equation}
and
\begin{equation}
{\rm ierfc}(\z)=\frac{1}{\sqrt{\pi}} e^{-\z^{2}}-\z {\rm erfc}(\z).
\end{equation}
This gives the approximate solution of $\phi$ as
\be
\phi \=\phi_0-\phi_{0} \sqrt{4 \lambda' t}\,\,
{\rm ierfc}(\z)-\kk{\phi_0 \sqrt{\v' \phi_0}}{(1-\phi_0)} \, t
e^{-\kk{(1-\phi_0) z}{\sqrt{\v' \phi_0}}},
\label{sol-1}
\ee
thus compaction essentially occurs in a boundary layer near the
bottom with a thickness of the order of $\sqrt{\v'}$.
The comparison \, of
this approximate solution  (\ref{sol-1}) \,
(dashed curves) with numerical solutions (solid
curves) is shown in Figure 1 for the values of $\v=0.01,  \, t=5$.
The  approximate solution  is accurate when $(\phi/\phi_0)^m \ll 1$ so that
$t \sim 1/m\sqrt{\v} \sim 5$, due to the fact that $\phi(z=0)=
1-O(\sqrt{\v} t)$. The agreement is clearly shown
in the figure.

\begin{figure}
{\includegraphics[width=9cm]{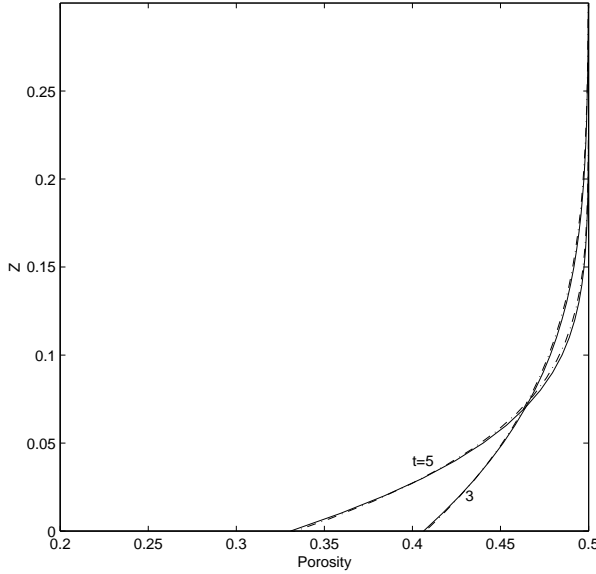}}
  \caption[]{ Comparison of numerical solutions (solid curves) with
          asymptotic solution (\ref{sol-1}) for $\v=0.01$
          at $t=3,5$. Where $Z=z/h(t)$ is the scaled height.  }
\end{figure}

\subsection{Fast Compaction ($\v \gg 1$)}

Either viscous or poroelastic fast compaction, is more complicated and
interesting in contrast to the simple   structure of the boundary layer for
slow compaction. Since $\v \gg 1$ and the highly nonlinear permeability
function $k=(\phi/\phi_0)^m, \,m=8$, the governing equations are also highly
nonlinear. However, we use these features and pursue asymptotic
analysis to seek appropriate asymptotic solutions.

\subsection{Poroelastic Compaction}

For the case $\v \gg 1$, we rewrite (\ref{equ-3}) as
\[  \pab{\phi}{t}+\v (\kk{\phi}{\phi_0})^m [-\pab{p}{z}
-(1-\phi)] \pab{\phi}{z} =-(1-\phi) p \]
\be
-\kk{\phi}{(1-\phi)} \{ \pab{p}{t}+ \v (\kk{\phi}{\phi_0})^m
[-\pab{p}{z}-(1-\phi) ] \pab{p}{z} \},
\ee
By using the perturbations
\be
\phi=\phi\0+\kk{1}{\v} \phi\1+ ..., \s p=p\0+\kk{1}{\v} p\1 + ...,
\ee
the leading order equation becomes
\be
\pab{\phi\0}{z}=-\kk{\phi\0}{(1-\phi\0)} \pab{p\0}{z},
\ee
whose integration gives
\be
p\0=\ln (\phi_0/\phi\0)-(\phi_0-\phi\0),
\ee
which is the Athy-type relation and is exactly the same
form as in Smith (1971) and Fowler and Yang (1998).
From (\ref{equ-2}) the equation of leading order is thus
\be
\kk{(1-\phi\0)}{\phi\0} \pab{\phi\0}{z}-(1-\phi\0)=0,
\ee
or
\be
\pab{\phi\0}{z}-\phi\0=0.
\ee
The boundary condition $\phi\0=\phi_0$ gives
\be
\phi\0=\phi_0 e^{-(h-z)},
\label{sol-2}
\ee
which decreases with depth $h-z$ exponentially. This solution is the
same as  the equilibrium solution in the poroelastic case, and thus
the top region of viscoelastic compaction is essentially
poroelastic and the viscous effect is only of secondary importance in
this region. However, as $\phi$ decreases, the term $\v (\phi/\phi_0)^m$
may become very small due to the large exponent $m=8$. The relation
$\v (\phi/\phi_0)^m=1$ defines a critical value of $\phi$ in the
transition region
\be
\phi^*=\phi_0 e^{-\kk{\ln \v}{m}}.
\ee
In fact, the above solutions are only valid when $\phi\0 > \phi^*$ and
$h-z<\Pi=(\ln \v)/m$.

\subsection{Transition Region}

Since $\phi \sim \phi^*$, we define
\be
\phi=\phi^* e^{\kk{\psi-\ln m}{m}}, \s
z=h-\Pi+\kk{\n-\ln m}{m},
\ee
\be
u^s=\kk{W}{m}, \s p=p^*-\kk{P}{m}, \label{equ-300}
\ee
where $p^*=\ln (\phi_0/\phi^*)-(\phi_0-\phi^*)$.
By changing variables to ($t, \n$) via $\p_t \ra \p_t-m \dot h \p_{\n}$,
$\p_z \ra m \p_{\n}$, and assuming $m \gg 1$,
we have the equation of leading order
\be
\dot h \phi^* \psi_{\n}+(1-\phi^*) W_{\n}=0,
\ee
\be
W=e^{\psi} [P_{\n}-(1-\phi^*)],
\ee
\be
\dot h \phi^* \psi_{\n}=\kk{\phi^* \dot h}{(1-\phi^*)} P_{\n}+
(1-\phi^*) p^*.
\label{equ-200}
\ee
Thus
\be
W=W^*-\kk{\dot h \phi^*}{(1-\phi^*)} \psi,
\label{equ-100}
\ee
\be
\psi_{\n}-1=\kk{(1-\phi^*) p^*}{\dot h \phi^*}
+(\psi_{\infty}-\psi) e^{-\psi},
\ee
\be
\psi_{\infty}=\kk{(1-\phi^*) W^*}{\dot h \phi^*},
\label{sol-3}
\ee
whose solution can be written as a quadrature. In the limit
 $\n \ra -\infty$, $P_{\n}
\ra 0$, equation (\ref{equ-200}) shows that the dominant term is
the viscous term $(1-\phi^*) p^*$ so that compaction
will gradually shift from viscoelastic to purely
viscous behavior. This has important geophysical implication for compaction
in sedimentary basins, since the purely viscous mechanism may be
responsible for overpressuring and mineralized seals in oil-reservoir
and hydrocarbon basins.
Furthermore, In order to determine $\dot h$, we require
(\ref{equ-100}) and (\ref{sol-3}) to match the solution
below the transition layer.

\subsection{Viscous Compaction}

In the region below the transition layer, the porosity $\phi< \phi^*$ is
usually very small, while the effective pressure $p$ is increasing and
$p \sim p^*=O(1)$. Rewriting (\ref{equ-3}) as
\be
\pab{u^s}{z}=-\kk{\phi}{(1-\phi)^2}[\pab{p}{t}+u^s \pab{p}{z}]- p,
\ee
From (\ref{equ-300}) and (\ref{equ-200}), we know that $p$ changes
slowly, which implies $\pab{p}{t} \sim \pab{p}{z} \ll 1$ or
$\phi (\pab{p}{t}+u^s \pab{p}{z}) \ll p$. We then have approximately
\be
p \= - \pab{u^s}{z},
\ee
which implies that compaction is now essentially purely viscous.
Thus we get
\begin{equation}
\pab{\phi}{t} =\pab{}{z}[(1-\phi) u^s],
\end{equation}
\begin{equation}
u^s=\v (\kk{\phi}{\phi_0})^m [\pcd{u^s}{z}-(1-\phi) ],
\end{equation}
which are the equations solved by Fowler and Yang (1999) when $\Xi=1$
for purely viscous compaction.
Following the same solution procedure given by Fowler and Yang (1999),
we can expect to get the same solutions. Thus, we only write down here
the solution for $\dot h$
\be
\dot h=\dot m_s [\kk{(1-\phi_0)}{(1-\phi^*)}+\kk{\phi^* \psi_{\infty}}
{m (1-\phi^*)}]-\kk{2 p^*}{\gamma m} \ln m,
\ee
where $\gamma=\kk{p^*(1-\phi^*)^2}{\dot m_s \phi^* (1-\phi_0)}$.
This essentially completes the solution procedure. Figure 2 shows the
comparison of numerical results with the above obtained asymptotic solutions
(\ref{sol-2}) and (\ref{sol-3}) in the poroelastic and transition region.

\begin{figure}
{\includegraphics[width=9cm]{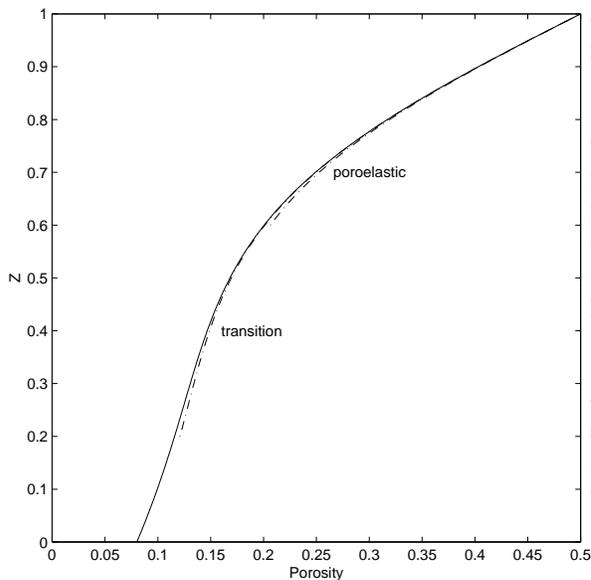}}
  \caption[]{Numerical results (solid curves)
          with $\v=100, \, t=10$.
          The asymptotic solutions (\ref{sol-2}) and (\ref{sol-3})
          are also plotted as a comparison (dashed).
          Profile in the top region is nearly exponential follows
          by a transition to pure viscous compaction where
          porosity is nearly uniform.}
\end{figure}

\section{Discussion}

The present model of viscoelastic flow and nonlinear compaction
in sedimentary basins uses a rheological relation which incorporates
 both poroelastic and viscous effects in the 1-D compacting frame.
Based on the frame invariance of irrotational feature of the 1-D flow,
a generalised viscoelastic compaction relation of Maxwell type
has been formulated.
The nondimensional model equations are mainly
controlled by one parameter $\v$, which is the ratio of the
hydraulic conductivity to the sedimentation rate.
Following a similar asymptotic analysis given
by Fowler and Yang (1998), we have been able to obtain the
approximate solutions for either slow compaction ($\v \ll 1$)
or fast compaction ($\v \gg 1$). The more realistic and yet
more interesting
case is when $\v \gg 1$, and the  solution gives
a nearly exponential profile of porosity versus depth,
which implies that compaction in the top region is essentially
poroelastic and its profile is virtually at equilibrium.

The numerical simulations and asymptotic analysis have shown
that porosity-depth profile is nearly exponential followed by a
a transition from a poroelastic to a viscoelastic  regime.
This is because the
large exponent $m$ in the permeability law $k
=(\phi/\phi_{0})^{m}$; even if $\lambda \gg 1$,
the product $\lambda  k$ may  become small at
sufficiently large depths. In this case, the
porosity profile consists of an upper part
near the surface where $\lambda  k \gg 1$ and equilibrium is
attained, and a lower part where $\lambda  k \ll 1$, and the
porosity is higher than equilibrium, which appears to
correspond accurately to numerical computations.
Below  this transition region, porosity is usually
uniformly small and compaction is essentially purely viscous.
From the definition of excess pore pressure $p_{\rm ex}=\int^{z}_{h}
[p+(1-\phi)] dz$, we know that the sudden switch from
poroelastic to viscous compaction means a quick decease of porosity $\phi$,
which leads to a sudden increase of $p_{\rm ex}$. Therefore,
the transition is often associated with a jump to a high pore
pressure and low permeability region where a mineralized
seal may be formed. This conclusion is consistent with the
earlier work (Birchwood and Turcotte, 1994). As viscous compaction
proceeds, porosity and  permeability may become so small that
fluid gets trapped below this region, and compaction virtually
stops.

Further work shall focus on more realistic and correct formulation
of rheology. In a recent work on pressure solution and its
application to some  field problems such as land subsidence
associated with fluid withdrawal from undercompacted aquifers,
Revil (1999) suggests a Voigt-type poro-visco-plastic rheological
behavior to characterize pressure solution and to applications to
some field problems including equilibrium and disequilibrium
compactions and subsidence. Naturally, more work is needed to
incorporate a Voigt-type rheology applied to compaction in addition
to  the  present Maxwell-type law.

{\it Acknowledgements}:
The author thanks the referees
for their very helpful comments and very instructive suggestions.
I also would like to thank Prof. Andrew C Fowler for his very helpful
direction on viscous compaction.

\end{document}